# Stabilizing Itinerant Electrons in a Corner-Sharing Kagomé Oxide $Nd_4Os_3ZnO_{14}$

Ryutaro Okuma[1], Yuita Fujisawa[2], Chia-Hsiu Hsu[3], Keita Kojima[1], Daisuke Nishio-Hamene[1], Kazuki Sumida[2], Akio Kimura[4,5,6], Akira Yasui[7], Kohei Yamagami[7], Jun-ichi Yamaura[1], Yoshihiko Okamoto[1]

[1]Institute for Solid State Physics, University of Tokyo, Kashiwa, Chiba 277-8581, Japan

[2]Research Institute for Synchrotron Radiation Science (HiSOR), Hiroshima University, Higashi-Hiroshima 739-0046, Japan.

[3]Quantum Materials Science Unit, Okinawa Institute of Science and Technology (OIST), Okinawa 904-0495, Japan.

[4]Graduate School of Advanced Science and Engineering, Hiroshima University, Higashi-hiroshima, Hiroshima 739-8526, Japan.

[5]International Institute for Sustainability with Knotted Chiral Meta Matter (WPI-SKCM[2]), Hiroshima University, Higashi-hiroshima, Hiroshima 739-8531, Japan.

[6]Research Institute for Semiconductor Engineering, Higashi-Hiroshima, Hiroshima 739-8527, Japan.

[7]Japan Synchrotron Radiation Research Institute (JASRI), Sayo, Hyogo, 679-5198 Japan.

**Abstract**: Kagomé oxides provide a fertile platform for exploring exotic electronic states arising from geometrical frustration and characteristic band topology. Here, we report the synthesis of a 5*d* transition-metal kagomé oxide, $Nd_4Os_3ZnO_{14}$, obtained via high-temperature, high-pressure hydrothermal synthesis. Single-crystal X-ray diffraction reveals a two-dimensional kagomé network formed by corner-sharing $OsO_6$ octahedra, with a nominal osmium valence of +4.67. In-plane resistivity and hard X-ray photoelectron spectroscopy measurements indicate that the semimetallic electronic structure at room temperature evolves into a semiconducting ground state upon cooling, accompanied by a pronounced enhancement of hole mobility. Magnetic susceptibility measurements demonstrate localized $Nd^{3+}$ moments without long-range magnetic order down to 2 K. The coexistence of a metallic kagomé plane, strong spin–orbit coupling inherent to 5*d* electrons, and rare-earth magnetism establishes $Nd_4Os_3ZnO_{14}$ as a promising platform for investigating correlated electron phenomena in kagomé oxides within the itinerant regime.

## INTRODUCTION

The kagomé lattice, composed of corner-sharing triangles, is a prototypical geometry for strong geometrical frustration. In antiferromagnetic systems, it has long been explored as a platform for realizing quantum spin liquid states. [1,2] In momentum space, tight binding model on the kagomé lattice hosts distinctive electronic features such as flat bands, linear band crossings, and van Hove singularities, providing an ideal setting for emergent electronic phenomena. [3-8]

Figure 1 summarizes representative transition-metal compounds featuring kagomé networks. In insulating kagomé-lattice antiferromagnets, the kagomé geometry typically arises from edge- or corner-sharing octahedra of transition-metal ions, as exemplified by copper-based minerals[9-17] and jarosites.[18] In contrast, kagomé-lattice compounds exhibiting metallic behavior adopt fundamentally different structural motifs. Materials with strong covalent bonding character are constructed from face-sharing octahedra with enhanced direct orbital overlap, [19,20] while intermetallic compounds realize kagomé lattices within close-packed metal frameworks.[21]

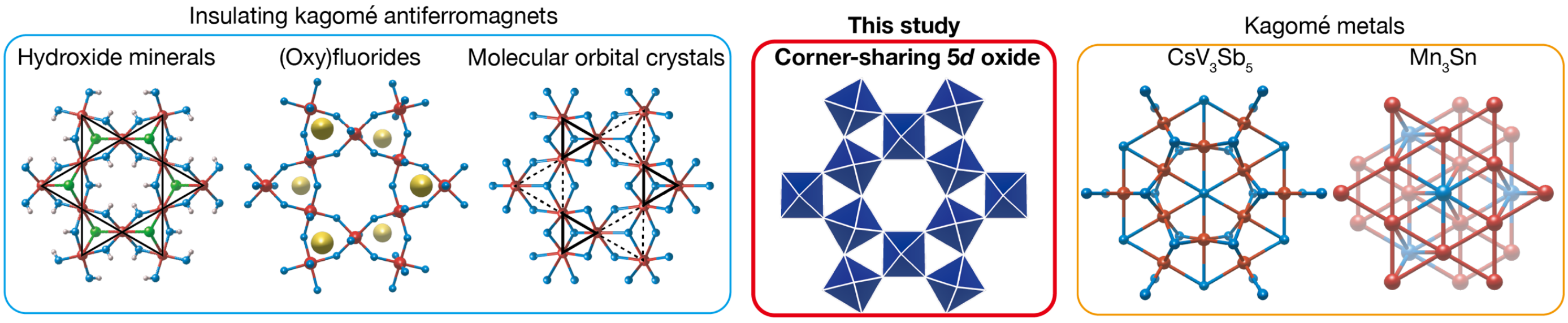

**Figure 1**. **Structural diversity of *d*-electron-based kagomé materials.** Red and light-blue spheres represent transition-metal ions and ligands, respectively. Copper hydroxides[4] consist of edge-sharing octahedral networks. Corner-sharing octahedral networks have so far been realized only in strongly ionic compounds, such as jarosite,[18] vanadium oxyfluorides[22] and copper/titanium fluorides.[23-26] In cluster compounds,[27-30] the edge-sharing octahedra are strongly distorted, resulting in molecular triangular units rather than an extended kagomé network. Metallic kagomé compounds, by contrast, typically feature face-sharing octahedra[19,20,31] or close-packed arrangements of metal atoms.[32] The corner-sharing 5*d* transition metal kagomé oxides investigated in this study therefore represent a missing structural class within the family of kagomé materials.

A central challenge has been the difficulty of accessing an intermediate regime that combines electron itinerancy with strong correlations. Motivated by the possibility of exotic superconductivity emerging from doped quantum spin liquids,[33] carrier doping has been attempted in insulating kagomé-lattice antiferromagnets; however, such efforts have consistently resulted in strong charge localization.[34-37] In edge-sharing kagomé networks with fractional electronic filling, electron itinerancy is suppressed by the formation of localized molecular orbitals.[27-30] As a result, kagomé-lattice oxides that combine strong electron correlations with itinerant charge degrees of freedom have remained elusive.

Very recently, kagomé-lattice oxynitrides predicted to exhibit metallic behavior were synthesized via ammonia-assisted reduction of layered molybdenum oxides.[38] Motivated by the potential of 4*d* and 5*d* transition metals to support correlated itinerant states, we targeted the synthesis of corner-sharing kagomé-lattice oxides composed of 5*d* ions. These ions naturally possess wider electronic bandwidths and are less susceptible to molecular-orbital–driven localization. Such kagomé geometries can, in principle, be realized as ordered substructures within crystallographic frameworks common to 5*d* oxides, including pyrochlore- and weberite-type structures.[39-42]

Through systematic materials exploration using hydrothermal synthesis, we have succeeded in growing single crystals of a novel transition-metal-based kagomé-lattice oxide that exhibits metallic transport and pronounced electronic instabilities. This material establishes a new platform for investigating correlated electron physics in kagomé oxide systems.

## EXPERIMENTAL SECTION

**Single crystal growth of $Nd_4Os_3ZnO_{14}$.** We initially targeted the synthesis of the Os-analogue of a tripod kagomé $Zn_2Nd_3Sb_3O_{14}$,[39-41] where $Sb^{5+}$, having comparable ionic radius as $Os^{5+}$, forms a regular kagomé net. Small single crystals were synthesized by hydrothermal reaction under high-temperature conditions. In a typical synthetic condition, 0.625 mmol of ZnO, 0.1 mmol of $Nd_2O_3$, 0.2 mmol of Os metal, 0.25 mmol of $KClO_4$, 0.1 g of KOH and 0.2 ml of de-ionized $H_2O$ was put in a 5 cm-long silver tube. Both ends of the silver tube was sealed by flame and put in a stainless autoclave, filled with water. The vessel is kept at 750˚C and 200 MPa for 48 hours. After the reaction, black hexagonal thin platelets of $Nd_4Os_3ZnO_{14}$ with maximum dimensions of ~$0.9 \times 0.3 \times 0.01$ $mm^3$ (as shown in Fig. 2) crystallized together with $Nd_3OsO_7$, ZnO, $Nd(OH)_3$, and $K_2OsO_2(OH)_4$. The crystals are rinsed with water to remove KOH and mechanically separated from other products and dried in air.

**Structural Characterization.** Single crystal X-ray diffraction (XRD) measurements were performed on a a Rigaku XtaLAB system equipped with a semiconductor detector (HyPix-6000) using a microfocused Mo *Kα* radiation source (λ = 0.71069 Å; Rigaku MicroMax-007). The X-ray data were collected and reduced using the CrysAlisPro software. Absorption corrections were made empirically by defining the crystal shape. As non-merohedral twinning was observed, a domain with 120˚ rotation about the $c^*$ axis was included to index the data. For the crystal structure determination, the initial structure is solved using the integrated intensity of one major domain and the refinement was performed against all the integrated intensity from the major domain with other domain contributions. The initial model of the crystal structures was obtained by SHELXT[43] and refined using the program Olex2[44] and SHELXL.[45] The details of the structural refinement are provided in the Supplementary Information. VESTA was used for the crystal structure visualization.[46] Elemental analysis was performed using a scanning electron microscope equipped with an energy-dispersive X-ray spectrometer, confirming the presence of Nd, Os, and Zn in the expected stoichiometric ratio.

**Magnetometry.** Magnetization measurements were performed using a Quantum Design MPMS3 system in fields up to 7 T and temperatures down to 2 K.

**Resistivity measurement.** Five probe resistivity measurements were performed using a Quantum Design Physical Property Measurement System in fields up to 10 T and temperatures down to 2 K. Conductive silver paste is attached to the five sides of a single crystal with dimensions of 0.05 mm x 0.5 mm x 0.7 mm, and cured at 300˚C in air to reduce contact resistance down to ~1 Ω.

**Hard X-ray photoelectron spectroscopy (HAXPES).** HAXPES was performed at BL09XU of SPring-8. The sample was cleaved in situ under ultra-high vacuum. A photon energy of 7940 eV was used to probe the bulk electronic structure. The Fermi level was carefully calibrated at each temperature using a polycrystalline Au reference. The energy resolution is approximately 180 meV.

**Electronic structure calculation.** First-principles calculations were carried out within the framework of density functional theory (DFT)[47] using the Vienna Ab initio Simulation Package.[48, 49] The interactions between ions and valence electrons were described by the projector augmented wave method.[50] The exchange–correlation energy was treated using the Perdew–Burke–Ernzerhof functional within the generalized gradient approximation.[51-53] A plane-wave kinetic energy cutoff of 500 eV was employed throughout the calculations. The structural parameters were fixed to the experimentally determined values throughout the calculations. The *C*-centered monoclinic cell was transformed to the primitive cell. The total energy convergence criterion for electronic self-consistency was set to $10^{-6}$ eV. Γ-centered Monkhorst–Pack k-point mesh of $6 \times 6 \times 4$ was used for Brillouin zone sampling, and spin–orbit coupling was included in all electronic structure calculations. The Nd-4*f* states were treated as core electrons.

## RESULTS

**Synthesis and Phase Formation of $Nd_4Os_3ZnO_{14}$.** Hydrothermal synthesis was employed to stabilize $Nd_4Os_3ZnO_{14}$ under controlled oxygen chemical potential and to promote continuous nucleation of crystallites suitable for physical property measurements. The average osmium valence was tuned through the amount of $KClO_4$, while KOH acted both as a mineralizer for ZnO and $Nd_2O_3$ and as a reactant to form the highly water-soluble $Os^{6+}$ species $K_2OsO_2(OH)_4$.

Phase formation was found to occur predominantly above 650 °C; no reaction between Nd and Os was observed below this temperature. After 1–2 days of reaction, plate-like black crystals of $Nd_4Os_3ZnO_{14}$ with lateral dimensions up to the sub-millimeter scale were obtained (Fig. 2). Substitution of Nd with Eu yielded isostructural crystals, whereas attempts to grow nonmagnetic analogues using Y or La were unsuccessful, likely due to the limited solubility of their oxides under the present conditions. In all cases, a large excess of ZnO was essential to stabilize the kagomé phase relative to competing Nd–Os oxides such as $Nd_3OsO_7$ and $Nd_2Os_2O_7$.

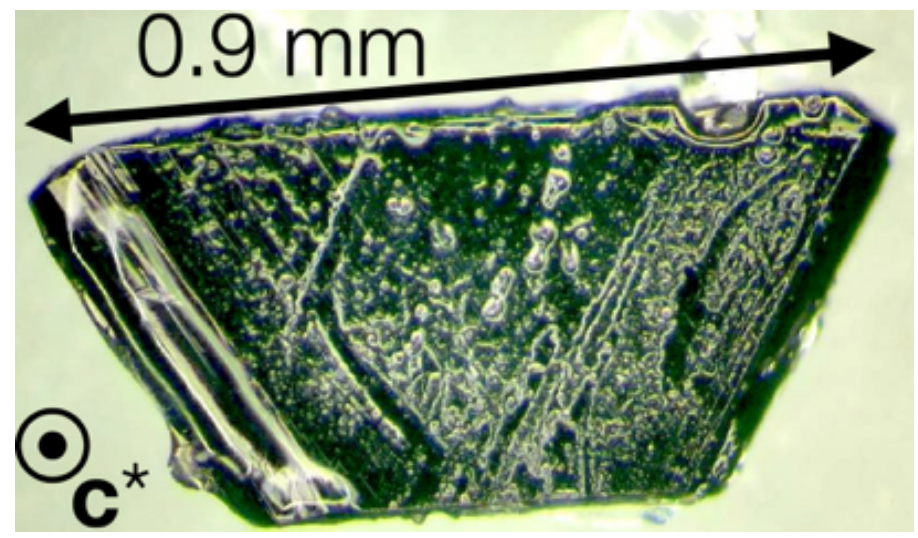

**Figure 2**. Optical microscope image of a single crystal of $Nd_4Os_3ZnO_{14}$ viewed from the $c^*$ direction.

**Crystal Structure and Kagome Lattice Formation.** Despite the hexagonal morphology of the crystals, single-crystal XRD revealed that $Nd_4Os_3ZnO_{14}$ crystallizes in a monoclinically distorted structure. As a result, all synthesized crystals exhibit twinning. The diffraction pattern can be indexed by two or three *C*-centered monoclinic domains related by 120° rotations about the monoclinic $c^*$ axis (Supplementary Information Section I). The non-merohedral twinning originates from the pseudo-trigonal metric of the unit cell rather than from a weak distortion of a higher-symmetry structure, as numerous reflections from different domains are clearly separated in reciprocal space.

The refined crystal structure (Figs. 3a–d, Table 1) consists of quasi-two-dimensional kagomé layers formed by corner-sharing $OsO_6$ octahedra, separated by triangular layers of Nd ions with coordination numbers of seven or eight. The structure is closely related to the insulating $Ta^{5+}$ kagomé oxide $LiBi_4Ta_3O_{14}$, except that the Li position could not be determined in the previous study.[54] Within the Os kagomé plane, the Zn1 atom is significantly displaced from the center of the hexagonal void, giving rise to a strongly distorted $ZnO_4$ tetrahedron. The coordination environment is characterized by two short out-of-plane Zn1–O7 bonds [1.870(7) Å] and two intermediate in-plane Zn1–O1 bonds [2.201(6) Å], resulting in a highly anisotropic 2+2 coordination geometry. Consistent with this distortion, the O–Zn1–O bond angles span a wide range from 75.0(7)° to 156.6(4)°. The Zn1 displacement occurs uniformly along the $\pm b$ direction and alternates between adjacent kagomé layers. Similar off-centered 2+2 coordination motifs have been reported in pyrochlore-related oxides such as $Mn_2Sb_2O_7$.[55]

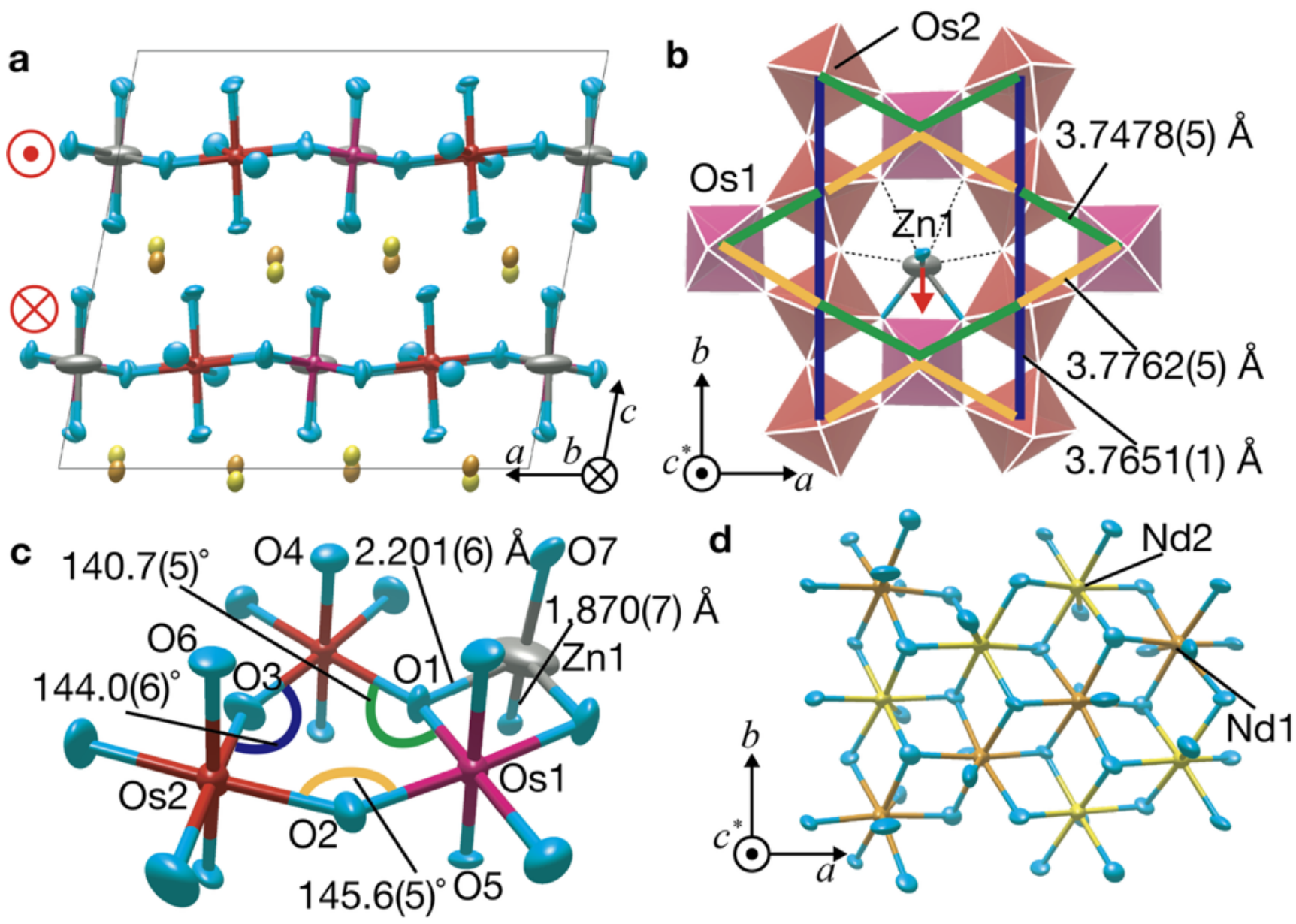

**Figure 3**. Crystal structures of $Nd_4Os_3ZnO_{14}$. Red/pink, sky-blue, grey, and yellow/orange spheres represent Os, O, Zn, and Nd atoms, respectively. Thermal ellipsoids are drawn at the 95% probability level. (a) Layered crystal structure viewed along the *b* axis. The black outline denotes the crystallographic unit cell. The direction of Zn distortion is shown in red ⊙ and ⊗. (b) Corner-sharing $OsO_6$ octahedral network viewed along the $c^*$ axis. Thick colored bonds indicate distinct Os–Os distances. The Zn1 site is displaced toward Os1, forming a 2+2 coordination with oxygen, while dashed bonds illustrate the remaining longer 2+6 coordination. (c) Local coordination environments of Os1, Os2, and Zn1. Inequivalent Os–O–Os bond angles are highlighted using different colors, following the scheme introduced in (b). (d) Triangular Nd layers formed by Nd1 and Nd2 sites with eight- and seven-fold coordination, respectively.

There are two crystallographically distinct Os sites: Os1, which shares oxygen ligands with Zn1, and Os2, which is located further away from the Zn-centered hexagon. Formal charge counting would suggest $Os^{4+}$ and $Os^{5+}$ for Os1 and Os2, respectively. However, the Os–O bond lengths (Table 2) differ only slightly between the two sites, yielding nearly identical bond-valence sums of +4.82 for Os1 and +4.76 for Os2, both close to the average valence expected from stoichiometry (+4.67). Within the kagomé plane, nearest-neighbor Os–Os distances range from 3.7478(5) to 3.7762(5) Å, while Os–O–Os bond angles span 140.7(5)° to 145.6(5)°, indicating only moderate distortions from an ideal kagomé geometry.

A residual electron-density peak of approximately +14 $e^{-}/Å^3$ remains near the Zn1 site in the disorder-free refinement model. Introducing split Zn and Nd positions substantially reduces the residual density and removes the checkCIF Alert B; however, the additional sites are approximately produced by pseudo-threefold operations associated with the pseudohexagonal twinning and possess nearly identical fractional $z$ coordinates. We therefore regard the split-site model primarily as a refinement of the diffraction intensity partitioning rather than definitive evidence for intrinsic positional disorder. Since the key structural parameters are essentially unchanged between the two models, the simpler disorder-free model is adopted in the main text. Details of the split-site refinement are provided in Supplementary Section III.

**Magnetic susceptibility dominated by $Nd^{3+}$.** The magnetic susceptibility measured with the magnetic field applied in the *ab* plane is shown in Fig. 4a. In the temperature range between 80 and 300 K, the susceptibility follows Curie–Weiss behavior. A Curie–Weiss fit yields an effective magnetic moment of 3.810(5) $\mu_B$ per Nd, a Weiss temperature of $T_{CW} = -52(1)$ K and a temperature-independent susceptibility of $\chi_0 = -4.1(1)\times10^{-4}$ $cm^3$•Nd-$mol^{-1}$. The extracted effective magnetic moment is in close agreement with the expected value of 3.62$\mu_B$ for $Nd^{3+}$ based on its Landé $g$ factor. No signatures of long-range magnetic ordering were observed down to 2 K, indicating that the magnetic response is dominated by localized $Nd^{3+}$ moments that remain paramagnetic within the measured temperature range.

**Metal-insulator-like anomaly.** The in-plane electrical resistivity of $Nd_4Os_3ZnO_{14}$ is shown in Fig. 4b. At temperatures above approximately 200 K, the resistivity exhibits metallic behavior ($d\rho/dT > 0$) with values on the order of mΩ·cm. Upon cooling, a clear anomaly appears at $T^* \approx 150$ K, defined by a cusp in $d\rho/dT$. Below this temperature, the resistivity increases rapidly, indicating a transition to an insulating-like state.

The Hall resistivity measurements reveal a complex evolution across $T^*$, including a crossover between electron-like and hole-like responses (Fig. 4c). Quantitative analysis based on a two-carrier model suggests a substantial reduction in the carrier density, decreasing from $n > 10^{19}$ $cm^{-3}$ at room temperature to approximately $10^{18}$ $cm^{-3}$ at the lowest temperature (see Supplementary Information). Concomitantly, a negative magnetoresistance is observed only within a narrow temperature window just below $T^*$, where the Hall response is electron-like (Figs. 4d). Upon further cooling well below $T^*$, electron scattering is progressively suppressed, and high mobility hole carriers dominate the transport signal, reaching $\mu_h > 800$ $cm^2$•$V^{-1}$•$s^{-1}$ at 2K. Overall, these results indicate a crossover from a low mobility correlated semimetal to a high mobility semiconducting regime across $T^*$.

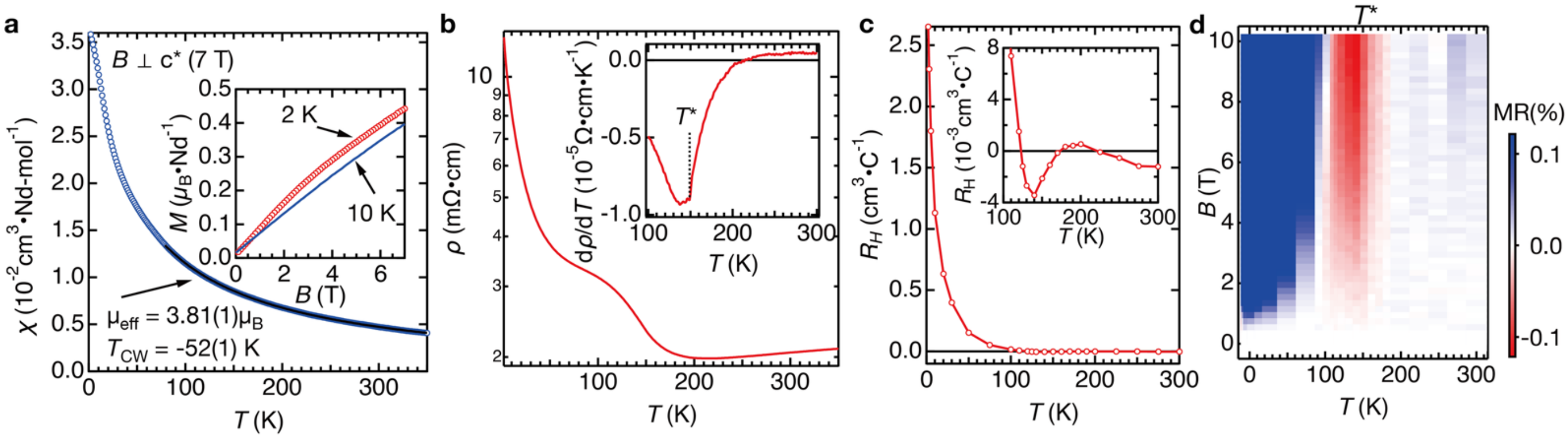

**Figure 4.** Physical properties of a twinned single crystal of $Nd_4Os_3ZnO_{14}$. (a) Magnetic susceptibility measured under an in-plane magnetic field of 7 T. The black solid line shows a Curie–Weiss fit performed in the temperature range 80–350 K. The inset shows magnetization process at 2 K (red open circles) and 10 K (blue line). (b) Temperature dependence of the in-plane electrical resistivity. The inset displays the temperature derivative of the resistivity, highlighting anomalies associated with the magnetic transition. (c) Temperature dependence of the Hall coefficient obtained from a linear fit to the Hall resistivity between 0 and 5 T. The inset shows a magnified view in the metallic regime. (d) Magnetic-field–temperature color map of the magnetoresistance. The color scale is shown on the right.

**Electronic Structure.** HAXPES measurements were carried out to probe the electronic structure of $Nd_4Os_3ZnO_{14}$ and were compared with DFT calculations. As shown in Fig. 5a, the spectral weight near the Fermi level is dominated by Os 5*d* states strongly hybridized with O 2*p* orbitals, while deeper-lying features originate from Zn 3*d*. A well-defined Fermi edge is observed at 200 K, consistent with the metallic transport behavior. Upon cooling, symmetrized HAXPES spectra reveal a suppression of spectral weight near the Fermi level (Fig. 5b), indicative of the opening of a gap or pseudogap across $T^*$.

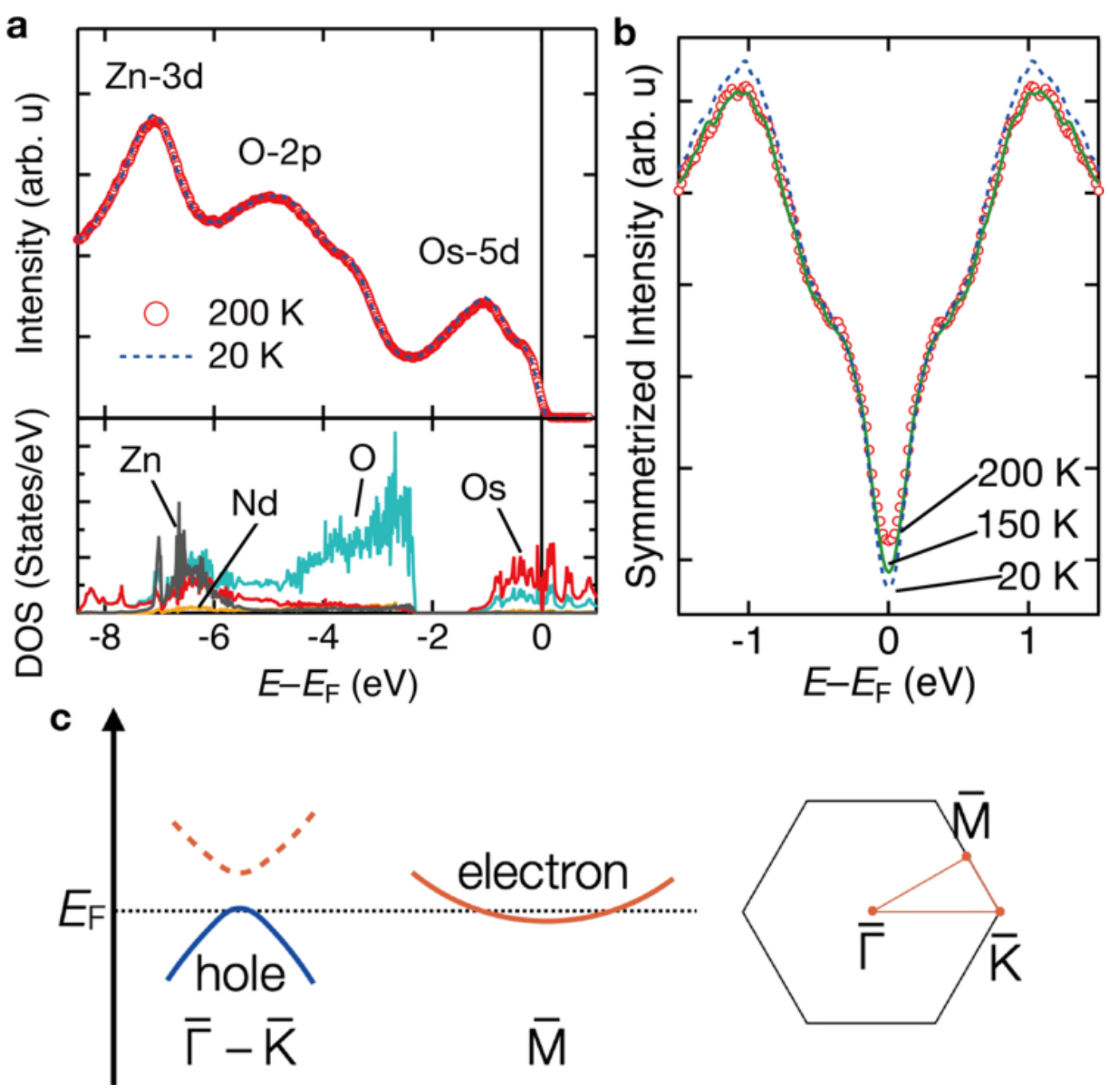

**Figure 5.** Electronic structure of $Nd_4Os_3ZnO_{14}$. (a) HAXPES spectra compared with the density of states (DOS) obtained from DFT calculations. (b) Symmetrized HAXPES spectra showing the suppression of spectral weight near the Fermi level across $T^*$. (c) Schematic illustration of the low energy electronic structure near the Fermi level. The electron and hole bands are shown by red and skyblue lines, respectively (left). The corresponding Brillouin zone is shown on the right. Although the crystal structure is monoclinic, the deviation from hexagonal symmetry is very small and $k_z$ dispersion is weak. For clarity, the Brillouin zone and high-symmetry directions are presented using a hexagonal representation, which captures the essential features of the electronic structure.

Band-structure calculations (Fig. 5c, Supplementary Information) yield a semimetallic electronic structure with small electron and hole pockets located along the Γ–K directions and near the M points, consistent with the nominal filling of ten electrons per three Os ions within a single layer. Notably, the bands near the Fermi level are relatively flat, particularly the electron-like band near the M point, which may naturally account for the weak temperature dependence of the transport coefficients observed experimentally.

## DISCUSSION

**Structural Origin of the Monoclinic Kagomé Framework.** A notable structural feature of $Nd_4Os_3ZnO_{14}$ is that its cation stacking sequence is intrinsically incompatible with threefold rotational symmetry, despite the apparent pseudohexagonal metric of the crystal. As illustrated in Fig. 6a, the cations occupy cubic close-packed positions, but the stacking registry of the kagomé layers places a vertex of the subsequent layer above the center of a triangle in the preceding layer. Consequently, the local threefold axis associated with the kagomé motif cannot become a crystallographic symmetry operation, making a rhombohedral description fundamentally impossible. This situation contrasts sharply with that of the related kagomé oxide $Nd_3Sb_3Zn_2O_{14}$, which adopts a rhombohedral structure despite a similar close-packed framework (Fig. 6b). In that compound, a different stacking phase preserves the threefold symmetry. Interestingly, the Zn site in $Nd_3Sb_3Zn_2O_{14}$ is also displaced from the center of the kagomé hexagon and statistically distributed around the threefold axis, suggesting an intrinsic tendency of $Zn^{2+}$ to favor a lower-coordination environment.

Importantly, the monoclinic symmetry of $Nd_4Os_3ZnO_{14}$ does not originate from the Zn displacement itself. Even with Zn constrained to the center of the kagomé hexagon, the stacking sequence remains monoclinic. Nevertheless, the pronounced off-centering of Zn toward an effective four-coordinate environment indicates that $Zn^{2+}$ energetically favors one of the symmetry-equivalent distortion directions. We therefore suggest that Zn off-centering further stabilizes the monoclinic stacking arrangement and removes the residual pseudo-threefold character that is preserved statistically in related tripod kagomé compounds. The origin of the different behavior between the osmate and antimonate systems remains unclear. One possible factor is the relatively low synthesis temperature of the hydrothermal growth process, which may favor a more ordered realization of the intrinsically monoclinic stacking sequence.

**Emergence of the Pseudo-Rhombohedral Metric.** The Nd sublattice itself forms an almost ideal cubic-close-packed triangular lattice, which naturally favors rhombohedral stacking. The pseudo-rhombohedral metric therefore originates primarily from the close-packed cation framework rather than from the kagomé network. The same close-packed stacking geometry that enforces monoclinic symmetry also gives rise to a pronounced pseudo-rhombohedral metric. To illustrate this, we define a transformed supercell by $\mathbf{a}_R = (\mathbf{a} - \mathbf{b})/2$, $\mathbf{b}_R = \mathbf{b}$, $\mathbf{c}_R = \mathbf{a} + 6\mathbf{c}$. The resulting unit cell is nearly rhombohedral, with $|\mathbf{a}_R| = |\mathbf{b}_R| = 7.52$ Å, $|\mathbf{c}_R| = 69.6$ Å, α' = 90.1°, β' = 90°, γ' = 120.1° (Fig. 6c). This relationship originates from the fact that neighboring kagomé layers are generated approximately by a translation of $\mathbf{b}/4 \pm \mathbf{c}/2$, which can be expressed as $(-2\mathbf{a}_R + 2\mathbf{b}_R \pm \mathbf{c}_R)/12$ in the transformed basis.

Consequently, four successive layers are required to generate a pseudo-rhombohedral centering operation. Although the local stacking sequence is incompatible with exact threefold symmetry, successive 2**c** and 4**c** translations progressively recover the rhombohedral stacking registry. The structure therefore exhibits a dual character: the local stacking geometry enforces monoclinic symmetry, whereas the long-period stacking sequence recovers a pseudo-rhombohedral registry on a larger length scale. This naturally explains both the strong pseudo-trigonal metric of the crystal and the ubiquitous formation of 120°-related twin domains.

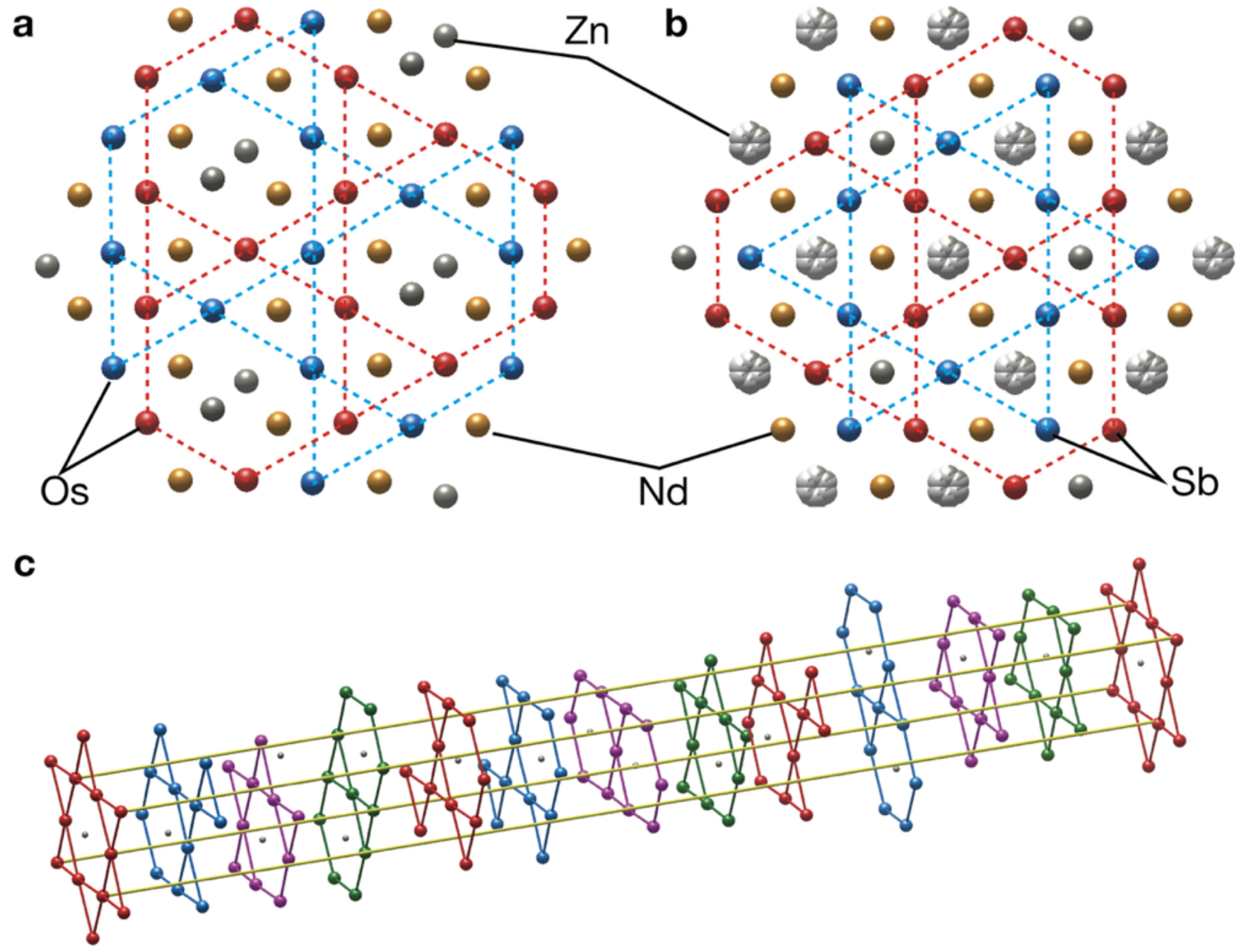

**Figure 6.** Cation stacking geometry in corner-sharing kagomé oxides. Dotted lines indicate the kagomé lattice formed by Os/Sb ions. (a) Monoclinic stacking geometry of the kagomé layers in $Nd_4Os_3ZnO_{14}$. (b) Rhombohedral stacking geometry of the cation framework in $Nd_3Sb_3Zn_2O_{14}$. (c) Recovery of a pseudo-rhombohedral stacking registry through successive 2**c**, 4**c**, and 6**c** translations in $Nd_4Os_3ZnO_{14}$. Kagomé layers related by the pseudo-rhombohedral centering translations are highlighted using the same color. The corresponding pseudo-rhombohedral supercell is outlined in yellow.

**Corner-Sharing Connectivity as a Chemical Route to Itinerant Kagomé Oxides.** The observation of metallic transport at high temperatures highlights the importance of octahedral connectivity in stabilizing itinerant electronic states in kagomé oxides. Most known transition-metal kagomé oxides are constructed from edge-sharing octahedra with metal–oxygen–metal bond angles close to 90°, a geometry that promotes localized electronic states and frequently leads to insulating ground states (Fig. 7a). In contrast, $Nd_4Os_3ZnO_{14}$ consists of corner-sharing $OsO_6$ octahedra with substantially larger Os–O–Os bond angles, placing it closer to the bonding regime found in pyrochlore and weberite-type oxides.

Consistent with this structural distinction, density functional theory calculations reveal electronic states near the Fermi level derived from extended Os 5*d* orbitals strongly hybridized with O 2*p* states, without signatures of molecular-orbital formation (Fig. 7b). This delocalized bonding framework naturally accounts for the observed itinerant transport despite the moderate local distortions associated with Zn displacement.

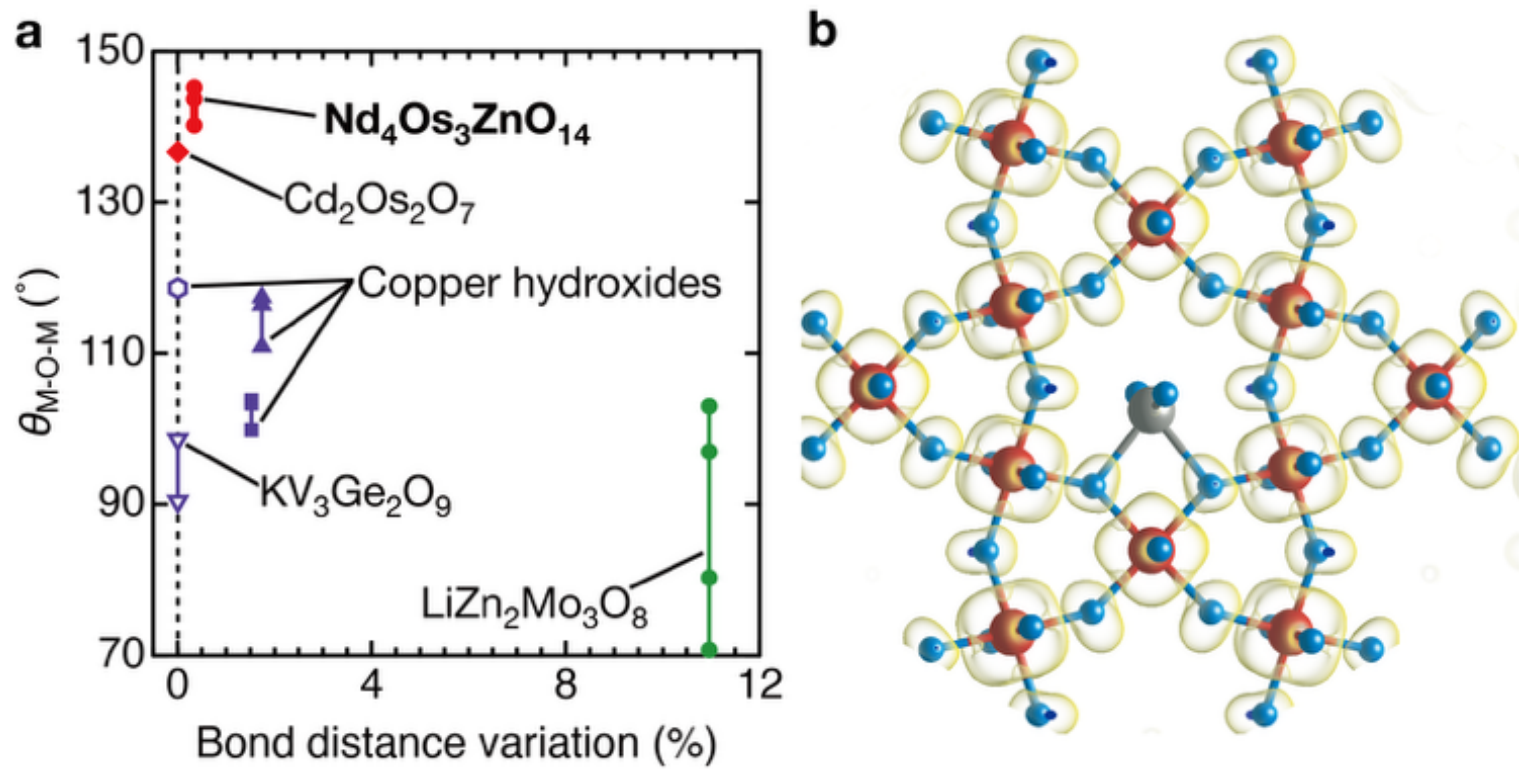

**Figure 7.** Absence of charge localization in $Nd_4Os_3ZnO_{14}$ at room temperature. (a) Comparison of metal–metal bond-length variation and metal–oxygen–metal bond-angle distribution for representative kagomé oxides. The normalized bond distance variation is defined as the standard deviation of the bond lengths divided by their mean value. Purple open hexagons, purple filled triangles, purple filled squares, purple open triangles, green filled circles, red filled squares, and red filled circles correspond to oxygen-metal-oxygen angle and normalized distribution of the metal-metal bond distance for herbertsmithite,[56] $Y_3Cu_9(OH)_{19}Cl_8$,[57] volborthite,[58] $KV_3Ge_2O_9$,[59] $LiZn_2Mo_3O_8$,[60] pyrochlore $Cd_2Os_2O_7$,[61] and $Nd_4Os_3ZnO_{14}$. For the copper hydroxides, bond angles

involving apical oxygens are not considered because of the pronounced Jahn-Teller distortion. (b) Charge density isosurface near the Fermi level calculated by DFT (–0.1 eV to 0 eV). The isosurface maximum is set to 0.001 $e^{-}$•$Å^{-3}$.

**Possible Origin of the Electronic Instability.** Having established the structural origin of the itinerant electronic state, we now consider the anomaly observed at $T^{*}$ ~ 150 K. The coexistence of mixed-valence osmium, geometrical frustration, and a semimetallic electronic structure suggests that the instability is electronic in origin rather than driven by a conventional structural phase transition. Indeed, synchrotron X-ray diffraction measurements below $T^{*}$ reveal no superlattice reflections or change in the structural parameters, indicating the absence of long-range structural symmetry breaking (Supplementary Information).

A natural chemical interpretation of the formal valence $Os^{4.67+}$ would be a mixed-valence state consisting of localized $Os^{4+}$ and $Os^{5+}$ ions. In such a scenario, sizeable local moments would be expected from the Os sublattice. However, the electronic structure calculations indicate a semimetallic state with only a small band overlap, and the overall agreement between the calculated density of states and the HAXPES spectra suggests that the low-energy electronic states are more naturally described in terms of itinerant bands shared across the Os kagomé network.

From this perspective, the anomaly at ($T^{*}$) may reflect a subtle instability of a nearly compensated semimetal rather than the formation of well-localized Os moments. In such a system, a variety of weak-coupling electronic instabilities may become relevant, including density-wave-like,[62] excitonic,[63,64] or other Fermi-surface instabilities in spin-orbit-coupled metals.[65] While the present data do not allow us to distinguish among these possibilities, the semimetallic electronic structure of $Nd_4Os_3ZnO_{14}$ places it in a regime where such weak-coupling instabilities may naturally occur. Further investigations using local probes such as muon spin spectroscopy, as well as element-selective techniques including resonant X-ray diffraction and resonant inelastic X-ray scattering, will be required to determine whether the instability originates from magnetic, charge, orbital, multipolar, or excitonic degrees of freedom.

**Implications for Correlated Kagomé Oxide Chemistry.** The present results establish corner-sharing 5*d* kagomé oxides as a distinct materials platform situated between insulating kagomé antiferromagnets and strongly covalent kagomé metals. Whereas edge-sharing kagomé networks tend toward localization through molecular-orbital formation, corner-sharing connectivity provides a chemically robust route toward itinerant electronic states while preserving the essential kagomé topology. Given the flexibility of this structural motif with respect to rare-earth ions, transition metals, and nonmagnetic spacer cations, the present work opens a pathway toward systematic exploration of correlated electronic phenomena in kagomé oxides. $Nd_4Os_3ZnO_{14}$ therefore represents not only a new kagomé compound but also an initial example of a broader class of metallic corner-sharing kagomé oxides.

## CONCLUSION

We have synthesized and characterized a novel kagomé oxide, $Nd_4Os_3ZnO_{14}$, featuring a two-dimensional network of corner-sharing $OsO_6$ octahedra stabilized by high-temperature, high-pressure hydrothermal synthesis. Single-crystal XRD reveals a kagomé lattice with mixed-valence osmium and moderate local distortions induced by Zn displacement, without significant charge disproportionation. $Nd_4Os_3ZnO_{14}$ exhibits metallic transport at high temperatures and a metal–insulator–like electronic instability upon cooling, while its magnetic response is dominated by localized $Nd^{3+}$ moments with no long-range order down to 2 K. These results demonstrate that corner-sharing connectivity in 5*d* kagomé oxides provides an effective chemical route to stabilizing itinerant electronic states under strong geometrical frustration. The present work establishes corner-sharing 5*d* kagomé oxides as a promising platform for exploring correlated electronic behavior beyond the insulating limit.

## ACKNOWLEDGEMENTS

We thank R. Koshoji, H. Ishikawa, and Z. Hiroi for fruitful discussions. We are grateful to N. Katayama and T. Kubo for their help with the synchrotron XRD measurements. Synchrotron XRD experiments were conducted at the BL02B1 of SPring-8, Hyogo, Japan (Proposals Nos. 2025A1631 and No. 2025B1930). HAXPES experiments were performed with the approval of the Japan Synchrotron Radiation Research Institute (Proposal No. 2025B1747). RO acknowledges the readCIF software provided by R. Coldea. RO acknowledges financial support from JSPS KAKENHI (Grant No. 26K17091), JST ASPIRE (Grant No. JPMJAP2314) and JST PRESTO (Grant No. JPMJPR2591).

**Table I**. Crystallographic data and the results of structural refinement for $Nd_4Os_3ZnO_{14}$ from single-crystal X-ray data. Atomic fractional coordinates, and equivalent isotropic displacement parameters (in units of $10^{-2}$Å$^2$) with estimated standard deviations in parentheses.

| Formula | $Nd_4Os_3ZnO_{14}$ |
|---|---|
| Crystal system | Monoclinic |
| Space group | $C2/c$ |
| $a$ (Å) | 13.0285(4) |
| $b$ (Å) | 7.5268(2) |
| $c$ (Å) | 11.8108(4) |
| $\beta$ (°) | 100.678(3) |
| $V$ (Å$^3$) | 1138.13(6) |
| $Z$ | 4 |
| Radiation | Mo $K\alpha$ ($\lambda$ = 0.71073 Å) |
| Temperature (K) | 293(2) |
| $\mu$ (mm$^{-1}$) | 53.391 |
| $\theta_{max}$ (°) | 38.743 |
| Index ranges | $-22 \leq h \leq 22$, $-13 \leq k \leq 13$, $-20 \leq l \leq 20$ |
| Independent reflections | 5962 |
| Reflections with $I > 2\sigma(I)$ | 5314 |
| Parameters refined | 102 |
| $F_{000}$ | 2440 |
| $R_1$ ($I > 2\sigma$) | 0.0493 |
| $wR_2$ (all data) | 0.1401 |
| Goodness-of-fit on $\|F\|^2$ | 1.044 |
| Absorption correction | Spherical, empirical (CrysAlisPro) |
| Crystal size ($\mu$m$^3$) | 120×60×20 |
| Twin fractions | 0.626(3)/0.374(3) |
| Largest peak/hole ($e^-$/Å$^3$) | 14.5/–5.2 |

| Atom | $x$ | $y$ | $z$ | $U_{eq}$ |
|---|---|---|---|---|
| Nd1 | 0.37539(4) | 0.87600(7) | 0.49379(5) | 0.97(1) |
| Nd2 | 0.38180(5) | 0.36960(6) | 0.53070(5) | 0.92(1) |
| Os1 | 1/2 | 0.63257(7) | 3/4 | 0.80(1) |
| Os2 | 0.25000(3) | 0.38118(5) | 0.75472(3) | 0.767(9) |
| Zn1 | 1/2 | 0.0614(4) | 3/4 | 2.90(6) |
| O1 | 0.3932(6) | 0.4439(10) | 0.7349(7) | 1.1(1) |
| O2 | 0.6048(6) | 0.8273(11) | 0.7751(7) | 1.2(1) |
| O3 | 0.2036(8) | 0.6303(10) | 0.7279(9) | 1.5(2) |
| O4 | 0.2117(7) | 0.3550(10) | 0.5861(7) | 1.0(1) |
| O5 | 0.5140(6) | 0.6344(9) | 0.9188(7) | 0.8(1) |
| O6 | 0.2844(6) | 0.3971(10) | 0.9257(6) | 0.9(1) |
| O7 | 0.5243(7) | 0.1125(9) | 0.9075(7) | 0.9(1) |

**Table II**. Bond lengths of the nearest Os-O and Zn-O at 293 K with estimated standard deviations in parentheses.

| *M*-O | *d* (Å) |
|---|---|
| Os1-O1 | 1.973(8) |
| Os1-O2 | 1.988(9) |
| Os1-O5 | 1.968(9) |
| Os2-O1 | 1.980(9) |
| Os2-O2 | 1.992(9) |
| Os2-O3 | 1.978(8), 1.982(8) |
| Os2-O4 | 1.971(9) |
| Os2-O6 | 1.989(7) |
| Zn1-O1 | 3.188(9) |
| Zn1-O2 | 2.215(9) |
| Zn1-O3 | 2.762(11) |
| Zn1-O7 | 1.868(9) |

# Supplementary Information for "Stabilizing Itinerant Electrons in a Corner-Sharing Kagomé Oxide $Nd_4Os_3ZnO_{14}$"

Ryutaro Okuma[1], Yuita Fujisawa[2], Chia-Hsiu Hsu[3], Keita Kojima[1], Daisuke Nishio-Hamane[1], Kazuki Sumida[2], Akio Kimura[4,5,6], Akira Yasui[7], Kohei Yamagami[7], Jun-ichi Yamaura[1], Yoshihiko Okamoto[1]

[1]Institute for Solid State Physics, University of Tokyo, Kashiwa, Chiba 277-8581, Japan

[2]Research Institute for Synchrotron Radiation Science (HiSOR), Hiroshima University, Higashi-Hiroshima 739-0046, Japan.

[3]Quantum Materials Science Unit, Okinawa Institute of Science and Technology (OIST), Okinawa 904-0495, Japan.

[4]Graduate School of Advanced Science and Engineering, Hiroshima University, Higashi-hiroshima, Hiroshima 739-8526, Japan.

[5]International Institute for Sustainability with Knotted Chiral Meta Matter (WPI-SKCM[2]), Hiroshima University, Higashi-hiroshima, Hiroshima 739-8526, Japan.

[6]Research Institute for Semiconductor Engineering, Higashi-Hiroshima, Hiroshima 739-8531, Japan.

[7]Japan Synchrotron Radiation Research Institute (JASRI), Sayo, Hyogo, 679-5198 Japan.

## 1. Non-merohedral twinning

The non-merohedral twinning observed in $Nd_4Os_3ZnO_{14}$ is described here. The twinning originates from the pseudo-rhombohedral metric symmetry that emerges after the superlattice transformation described in the main text. Representative diffraction patterns are shown in Supplementary Figs. 1a-b. In addition to integer reflections predicted for a single domain structure (Supplementary Figs. 1c-d), additional reflections are observed at positions corresponding to half-integer indices along the $c^*$ direction. These extra reflections can be reproduced by applying ±120˚ rotation about the $c^*$ axis, as illustrated in Supplementary Figs. 1e-h.

The origin of the half-integer reflections at specific $hkl$ reflection can be understood as follows. The corresponding twin law $T_{\pm 120^\circ}$, which maps a reciprocal lattice vector $(h, k, l)^T$ of a minor domain to $(h', k', l')^T = T_{\pm 120^\circ}(h, k, l)^T$ in the reference frame of the major domain, is given by

$$T_{\pm 120^\circ} = \begin{pmatrix} -\frac{1}{2} & \frac{\sqrt{3}a}{2b} & 0 \\ \pm\frac{\sqrt{3}b}{2a} & -\frac{1}{2} & 0 \\ -\frac{3c\cos\beta}{2a} & \pm\frac{\sqrt{3}c\cos\beta}{2b} & 1 \end{pmatrix}.$$

Using the metric relations $a/b \sim \sqrt{3}$ and $c\cos(\beta)/a = -1/6$, an $(h, k, l)$ reflection from the +120˚-rotated domain is mapped to

$$\left(\frac{-h+3k}{2}, -\frac{h+k}{2}, l+\frac{h-k}{4}\right)$$

in the reciprocal lattice of the major domain. Such reflections overlap with those of the major domain when $(h-k)/2$ is even, but are separable when $(h-k)/2$ is odd. When both +120˚ and –120˚ twin domains are present, reflections arising purely from the major domain occur only when both $(h+k)/2$ and $(h-k)/2$ are odd. The simulations show that the measured crystal predominantly contains the +120˚-rotated twin domain, while the other –120˚-domain is present only in a much smaller fraction.

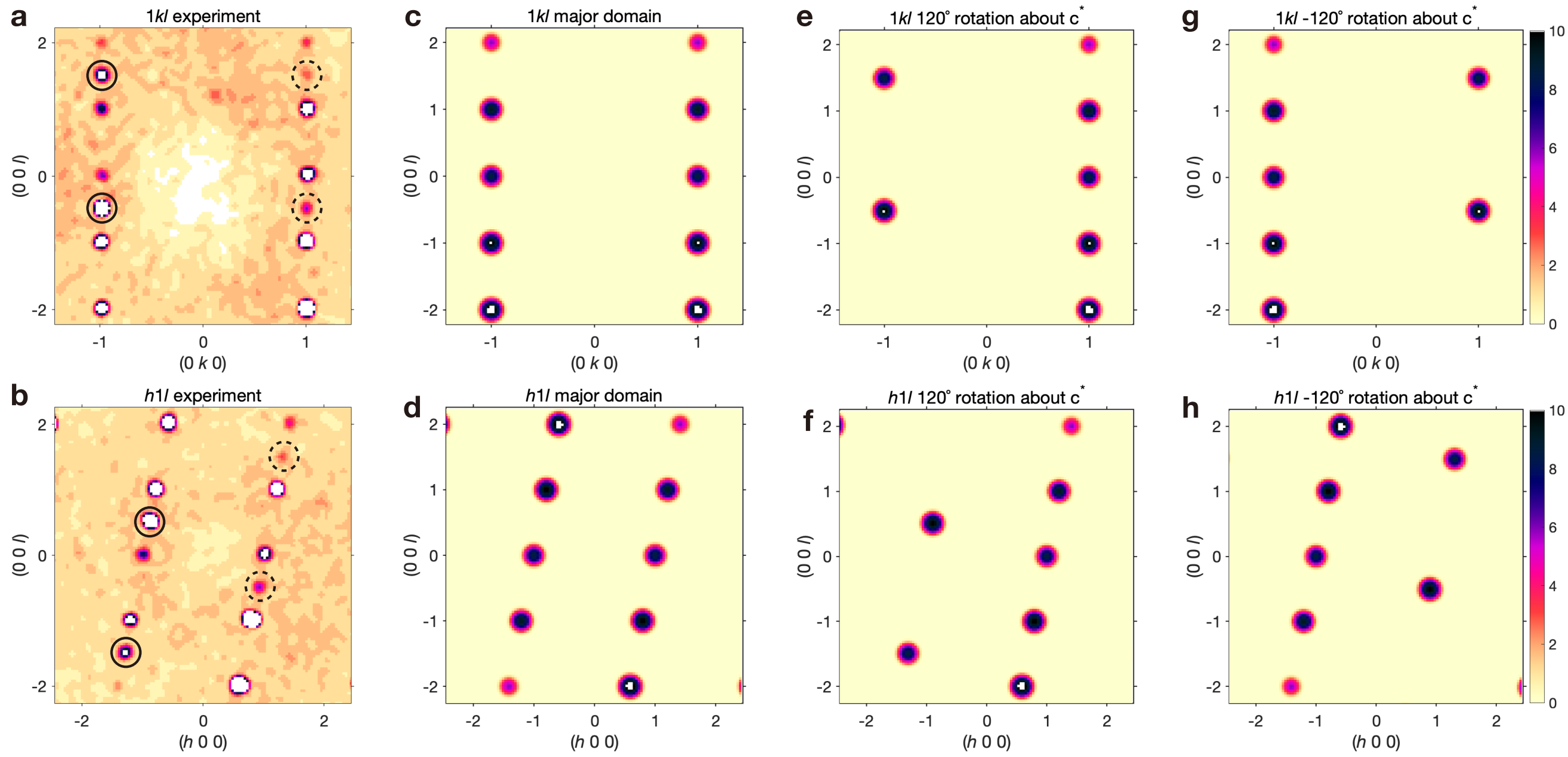

**Supplementary Figure 1. Twinning in $Nd_4Os_3ZnO_{14}$ crystal.** Reconstructed diffraction patterns in the $1kl$ plane (a) and $h1l$ plane (b). The corresponding simulated diffraction patterns from the major domain and 120˚ rotated domain and –120˚ rotated domain are shown in (c-h). Thick and dashed circles in (a) and (b) represent the reflections primarily from +120˚ and –120˚ rotated domains, respectively. The intensity scale is shown in the right.

## 2. Crystal structure refinement on the disorder free model

Here we describe how contributions from minor twin domains are treated in the refinement of the diffraction intensities. We first index the observed peaks using a single-domain unit cell while excluding non-integer reflections; this model is referred to as **Model: 1-domain#1**. A comparison with alternative refinement models that explicitly include the effects of twinning is summarized in Supplementary Table 1. The refined fractional atomic coordinates and atomic displacement parameters (ADPs) obtained from this model are nearly identical to those listed in Table 1 of the main text. Nevertheless, a substantially large residual electron density $\Delta\rho$ remains in the vicinity of the Zn1 site.
Next, reflections from the +120˚-rotated twin domain were included and the refinement was performed using the HKLF5 format (**Model: 2-domain#1**). The twin fraction of the minor domain was estimated to be approximately 37%. While the refined structure is essentially identical to that obtained in **1-domain#1** and the residual electron density near Zn is reduced the refinement yields higher $R_1$ and $wR_2$. We then excluded the non-integer peaks and refined the remaining intensities within the HKLF5 framework (**Model: 2-domain#2**). The refinement converged to the best agreement factors, with $R_1$ = 4.93%, $wR_2$ = 14.01% and goodness-of-fit of 1.044, together with a significantly reduced $\Delta\rho$ of 14.5/–5.2 e⁻/Å$^3$. This model is adopted in the main text and the corresponding structural parameters are listed in Table 1. For comparison, a refinement using only the major-domain reflections and excluding all contributions from the +120˚ domain (**Model: 1-domain#2**) yields the smallest $\Delta\rho$ of 8.5/–6.4 e⁻/Å$^3$ and the second-best agreement factors, as summarized in Supplementary Table 2. A refinement including reflections from the major domain as well as both ±120˚ twin domains (**Model: 3-domain**) results in the poorest agreement factors. This is most likely due to the extensive overlap of reflections from different domains, as discussed in the previous section.
Across all refinement models considered here, the refined structural parameters are largely consistent, with only marginal differences observed, except for the ADPs of certain oxygen atoms. The most prominent remaining issue is $\Delta\rho$ peak near Zn1. We speculate that the displacement of Zn1 may differ slightly between two domains, and that averaging over multiple domains leads to artificially enhanced ADPs and $\Delta\rho$. Such behavior could arise if the displacement originates from a high temperature phase transition, which will be investigated in future work.

**Supplementary Table 1. Comparison of various refinement models in the non-merohedrally twinned data.**

| Model | $N_{peak}$ | $R_1(I > 2\sigma)$ | $wR_{2,all}$ | GoF on $\|F\|^2$ | Twin fractions | Peak/hole (e⁻/Å$^3$) |
|---|---|---|---|---|---|---|
| 1-domain #1 | 3220 | 6.92 | 18.08 | 1.170 | – | 38.6/–9.1 |
| 2-domain #1 | 8409 | 7.16 | 22.57 | 1.052 | 0.628(3), 0.372(3) | 20.2/–9.4 |
| 2-domain #2 | 5962 | 4.93 | 14.01 | 1.044 | 0.626(3), 0.374(3) | 14.5/–5.2 |
| 1-domain #2 | 2706 | 5.84 | 15.89 | 1.113 | – | 8.5/–6.4 |
| 3-domain | 10440 | 8.26 | 29.82 | 1.184 | 0.549(5), 0.337(4), 0.114(1) | 36.4/–26.3 |

## 3. Crystal structure refinement on the split-site model

Here we describe a further attempt to account for the residual electron density remaining in the disorder-free refinement. We introduced split positions for atoms located near the major residual $Q$ peaks. For the Zn site (Supplementary Figure 2a), additional positions were generated by applying approximate threefold rotations about the center of the kagomé hexagon, motivated by the pseudo-trigonal symmetry associated with the twinning. This procedure yielded two additional Zn sites. Zn2 is displaced toward one of the O2 atoms, whereas Zn3 is located closer to the center of the hexagon (Supplementary Figure 2b). The refined occupancies correspond to Zn1 : Zn2 : Zn3 = 0.51 : 0.40 : 0.09. Importantly, all three Zn sites retain a pronounced tendency toward effective fourfold coordination. Even for Zn3, the in-plane Zn–O bond-length distribution clearly exhibits a 2+2 coordination pattern, supporting the picture of Zn off-centering within the kagomé hexagon. The remaining residual electron density near the Nd2 site can similarly be modeled by introducing an additional Nd position generated by an approximate threefold rotation of Nd1 (Supplementary Figure 2c). The refined occupancy ratio of the two Nd sites is 0.92 : 0.08. The split Zn and Nd sites possess nearly identical fractional $z$ coordinates. We therefore consider it unlikely that the apparent positional disorder directly reflects the intrinsic structure of a single-domain crystal. Instead, the split positions may partially originate from imperfect partitioning of diffraction intensities among overlapping reflections from the pseudohexagonal twin domains.

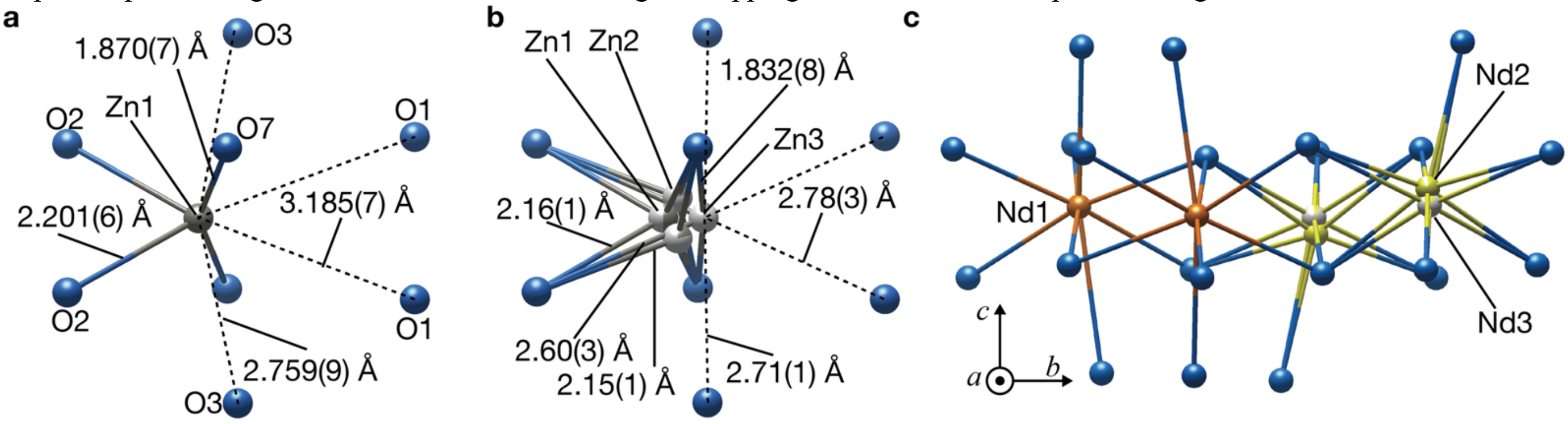

**Supplementary Figure 2. Split-site model for Zn and Nd atoms.** (a) Local coordination environment of the Zn1 site in the disorder-free model (used in the main text). (b) Local environments of the three refined Zn positions in the split-site model. (c) Split-site model for the Nd position.

Interestingly, when reflections from all three twin domains are included (3-domain model in Supplementary Table 2), the refined occupancies of the split Zn sites (0.50 : 0.36 : 0.14) become comparable to the independently refined twin fractions (0.55 : 0.34 : 0.11). This correspondence suggests that the split-site model may partially absorb residual contributions associated with the pseudo-trigonal twinning. However, the full three-domain refinement yields substantially larger residual electron densities because of severe reflection overlap, preventing a reliable quantitative analysis. We therefore do not regard the split positions as definitive evidence for intrinsic positional disorder. Instead, the split-site refinement likely captures a combination of the intrinsic Zn off-centering tendency and residual effects arising from the pseudohexagonal twin domains.

Nevertheless, the refined fractional coordinates of the remaining atoms are essentially unchanged upon introducing the split-site model. The principal structural features, including the Zn off-centering and the resulting effective 2+2 coordination environment, are therefore robust against the choice of refinement model. For this reason, the simpler disorder-free model is adopted for the structural discussion in the main text.

**Supplementary Table II**. Comparison with the disorder-free model (presented in the main text Table I) and split-site model. The refined peaks contain only integer position of the major domain, which were used in 2-domain #2. (Equivalent) isotropic displacement parameters are given in units of $10^{-2}$Å$^2$.

| | disorder-free model | | | | split-site model | | | | |
|---|---|---|---|---|---|---|---|---|---|
| $R_1$ ($I > 2\sigma$) | 0.0493 | | | | 0.0473 | | | | |
| $wR_2$ (all data) | 0.1401 | | | | 0.1345 | | | | |
| Goodness-of-fit on $\|F\|^2$ | 1.044 | | | | 1.049 | | | | |
| Largest peak/hole ($e^-$/Å$^3$) | 14.5/–5.2 | | | | 6.6/–5.5 | | | | |
| Atom | $x$ | $y$ | $z$ | $U_{eq}$ | $x$ | $y$ | $z$ | $U_{iso}/U_{eq}$ | Occ |
| Nd1 | 0.37539(4) | 0.87600(7) | 0.49379(5) | 0.97(1) | 0.37537(4) | 0.87598(7) | 0.49379(5) | 0.99(1) | 1 |
| Nd2 | 0.38180(5) | 0.36960(6) | 0.53070(5) | 0.92(1) | 0.38192(6) | 0.36937(9) | 0.5319(2) | 0.80(1) | 0.918(9) |
| Nd3 | – | – | – | – | 0.3792(7) | 0.3734(11) | 0.5085(18) | = Nd2 | 0.082(9) |
| Os1 | 1/2 | 0.63257(7) | 3/4 | 0.80(1) | 1/2 | 0.63257(6) | 3/4 | 0.82(1) | 1 |
| Os2 | 0.25000(3) | 0.38118(5) | 0.75472(3) | 0.767(9) | 0.24997(3) | 0.38099(5) | 0.75472(3) | 0.796(9) | 1 |
| Zn1 | 1/2 | 0.0614(4) | 3/4 | 2.90(6) | 1/2 | 0.0515(9) | 3/4 | 0.86(6) | 0.50(2) |
| Zn2 | – | – | – | – | 0.524(1) | 0.077(1) | 0.7493(5) | = Zn1 | 0.21(1) |
| Zn3 | – | – | – | – | 1/2 | 0.123(4) | 3/4 | = Zn1 | 0.088(9) |
| O1 | 0.3932(6) | 0.4439(10) | 0.7349(7) | 1.1(1) | 0.3931(6) | 0.4439(10) | 0.7348(7) | 1.1(1) | 1 |
| O2 | 0.6048(6) | 0.8273(11) | 0.7751(7) | 1.2(1) | 0.6048(6) | 0.8273(11) | 0.7753(7) | 1.2(1) | 1 |
| O3 | 0.2036(8) | 0.6303(10) | 0.7279(9) | 1.5(2) | 0.2037(7) | 0.6299(9) | 0.7281(9) | 1.5(2) | 1 |
| O4 | 0.2117(7) | 0.3550(10) | 0.5861(7) | 1.0(1) | 0.2117(6) | 0.3554(10) | 0.5861(6) | 0.9(1) | 1 |
| O5 | 0.5140(6) | 0.6344(9) | 0.9188(7) | 0.8(1) | 0.5141(6) | 0.6343(9) | 0.9190(6) | 0.9(1) | 1 |
| O6 | 0.2844(6) | 0.3971(10) | 0.9257(6) | 0.9(1) | 0.2845(6) | 0.3968(10) | 0.9256(6) | 0.9(1) | 1 |
| O7 | 0.5243(7) | 0.1125(9) | 0.9075(7) | 0.9(1) | 0.5246(7) | 1.1128(9) | 0.9077(6) | 1.0(1) | 1 |

## 4. Synchrotron X-ray diffraction result

To examine whether a structural change accompanies the anomaly in electrical resistivity observed around 150 K, we performed single-crystal synchrotron X-ray diffraction experiments at the BL02B1 beamline of SPring-8. The incident X-ray energy was set to 50 keV, and the sample temperature was controlled using an $N_2$ gas blowing system.

Supplementary Figures 2a and 2b show the reconstructed $h0l$ planes at 300 K and 100 K, respectively. A direct comparison of the diffraction patterns at these temperatures reveals that all observed Bragg peaks are identical, with no additional diffraction peaks appearing at low temperature. Furthermore, no peak splitting was detected. These observations indicate that neither the unit cell nor the extinction conditions change across the resistivity anomaly at 150 K. To further examine the possibility of symmetry lowering, synchrotron single-crystal X-ray diffraction datasets collected at 300 K and 100 K were independently refined using the same monoclinic $C2/c$ structural model. The resulting structural parameters are summarized in Supplementary Tables III and IV. Although the synchrotron measurements were optimized for detecting weak superlattice reflections and peak splitting rather than for high-precision structure determination, both datasets are satisfactorily described by the same $C2/c$ structural model. No systematic changes in the refined atomic positions or lattice symmetry were observed upon cooling. These results provide no evidence for a structural phase transition or symmetry reduction from $C2/c$ to $Cc$ down to 100 K.

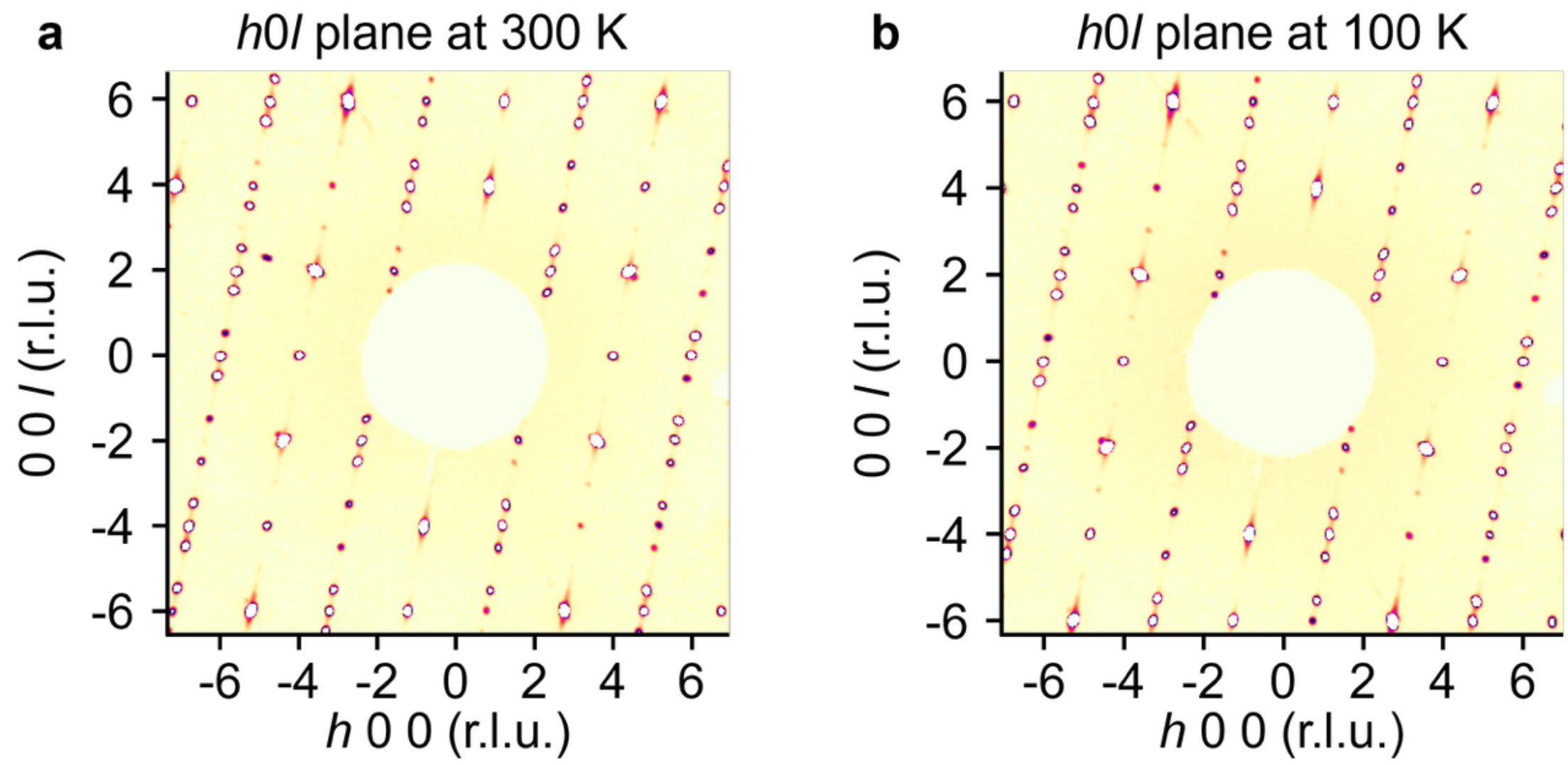

**Supplementary Figure 3. Absence of superstructure peaks in the synchrotron X-ray diffraction.** Reconstructed diffraction patterns in the *h*0*l* plane at 300 K (a) and 100 K (b).

**Supplementary Table III**. Crystallographic data and structural refinement parameters for $Nd_4Os_3ZnO_{14}$ determined from single-crystal synchrotron X-ray diffraction measurements at 300 and 100 K.

| Parameter | 300 K | 100 K |
|---|---|---|
| Formula | $Nd_4Os_3ZnO_{14}$ | $Nd_4Os_3ZnO_{14}$ |
| Crystal system | Monoclinic | Monoclinic |
| Space group | *C*2/*c* | *C*2/*c* |
| *a* (Å) | 13.0395(3) | 13.0299(3) |
| *b* (Å) | 7.5321(2) | 7.5254(2) |
| *c* (Å) | 11.8274(3) | 11.8093(3) |
| $\beta$ (˚) | 100.790(2) | 100.773(2) |
| *V* (Å$^3$) | 1141.09(5) | 1137.55(5) |
| *Z* | 4 | 4 |
| λ (Å) | 0.3093 | 0.3093 |
| Temperature (K) | 300 | 100 |
| Independent reflections | 1591 | 1583 |
| Reflections with $I > 3\sigma(I)$ | 1517 | 1510 |
| Parameters refined | 80 | 79 |
| $R(\|F\|)$ $[I > 3\sigma(I)]$ | 0.1092 | 0.1192 |
| $wR(\|F\|)$ | 0.1244 | 0.1330 |
| GOF | 9.21 | 9.55 |

**Supplementary Table IV.** Fractional atomic coordinates and isotropic displacement parameters (in units of $10^{-2}$Å$^2$) for $Nd_4Os_3ZnO_{14}$ obtained from the refinement of single-crystal synchrotron X-ray diffraction data at 300 and 100 K. Estimated standard deviations are given in parentheses.

| Atom | *x* (300 K) | *x* (100 K) | *y* (300 K) | *y* (100 K) | *z* (300 K) | *z* (100 K) | $U_{iso}$ (300 K) | $U_{iso}$ (100 K) |
|---|---|---|---|---|---|---|---|---|
| Os1 | 1/2 | 1/2 | 0.86712(11) | 0.86733(13) | 3/4 | 3/4 | 0.38(4) | 0.08(4) |
| Os2 | 0.24998(5) | 0.24997(6) | 0.61896(7) | 0.61847(8) | 0.74533(7) | 0.74523(8) | 0.54(4) | 0.30(4) |
| Nd1 | 0.37518(8) | 0.37497(9) | 0.62374(11) | 0.62397(13) | 0.49334(9) | 0.49276(10) | 0.60(4) | 0.25(4) |
| Nd2 | 0.38172(9) | 0.38172(10) | 0.13037(11) | 0.13058(13) | 0.53063(12) | 0.53106(13) | 0.66(4) | 0.37(4) |
| Zn1 | 1/2 | 1/2 | 0.4376(6) | 0.4418(6) | 3/4 | 3/4 | 2.77(9) | 1.61(9) |
| O1 | 0.3961(10) | 0.3971(11) | 0.6737(18) | 0.673(2) | 0.7245(12) | 0.7241(13) | 1.1(3) | 0.7(3) |
| O2 | 0.1068(9) | 0.1070(10) | 0.5548(15) | 0.5547(18) | 0.7652(10) | 0.7649(11) | 0.7(2) | 0.4(3) |
| O3 | 0.2050(11) | 0.2049(12) | 0.8693(14) | 0.8695(17) | 0.7287(13) | 0.7294(14) | 1.2(3) | 0.9(3) |
| O4 | 0.2859(10) | 0.2862(12) | 0.6447(14) | 0.6453(17) | 0.9125(12) | 0.9128(14) | 0.4(3) | 0.13 |
| O5 | 0.2176(9) | 0.2170(11) | 0.6036(16) | 0.6027(18) | 0.5742(12) | 0.5750(13) | 0.4(2) | 0.1(3) |
| O6 | 0.5137(10) | 0.5142(12) | 0.8662(13) | 0.8662(16) | 0.9187(12) | 0.9188(14) | 0.2(3) | 0.3(3) |
| O7 | 0.4751(9) | 0.4747(11) | 0.3873(13) | 0.3873(16) | 0.5918(12) | 0.5913(14) | 0.1(2) | 0.1(3) |

## 5. Two carrier fit of the resistivity

The experimentally measured longitudinal and Hall resistivities exhibit a pronounced temperature dependence. To extract information on the underlying electronic structure, the magnetic-field dependence of the resistivity tensor was analyzed using a standard two-carrier model, described by

$$\sigma_{xx} = q\left(\frac{n_e\mu_e}{1+(\mu_e B)^2} + \frac{n_h\mu_h}{1+(\mu_h B)^2}\right), \sigma_{xy} = qB\left(\frac{n_h\mu_h^2}{1+(\mu_h B)^2} - \frac{n_e\mu_e^2}{1+(\mu_e B)^2}\right)$$

$$\rho_{xx} = \frac{\sigma_{xx}}{\sigma_{xx}^2+\sigma_{xy}^2}, \rho_{yx} = \frac{\sigma_{xy}}{\sigma_{xx}^2+\sigma_{xy}^2}$$

Here $\sigma_{xx}$ and $\sigma_{xy}$ are the longitudinal and Hall conductivities, and $\rho_{xx}$ and $\rho_{yx}$ are the corresponding resistivities. $q = 1.6\times10^{-19}$ C is elementary charge, $B$ is external magnetic field, and $n_\mathrm{i}$ and $\mu_i$ ($i$ = e, h) denote the carrier density and mobility of electrons and holes, respectively. In the low field regime, the magnetoresistance and Hall resistivity reduce to MR $\propto B^2 n_\mathrm{e} n_\mathrm{h}(\mu_\mathrm{e}+\mu_\mathrm{h})^2/(n_\mathrm{e}\mu_\mathrm{e}+ n_\mathrm{h}\mu_\mathrm{h})^2$, $\rho_{yx} \propto (n_\mathrm{h}\mu_\mathrm{h}^2 - n_\mathrm{e}\mu_\mathrm{e}^2)/q(n_\mathrm{e}\mu_\mathrm{e}+ n_\mathrm{h}\mu_\mathrm{h})^2$. In this regime, where the magnetoresistance is quadratic in $B$ and the Hall resistivity is linear, the four parameters, $n_\mathrm{i}$ and $\mu_\mathrm{i}$ cannot be uniquely determined from transport data alone.

Motivated by the chemical formula and the band structure calculations, which suggest a compensated semimetallic state, we therefore first performed the fitting under the assumption of equal electron and hole densities ($n_\mathrm{e} = n_\mathrm{h}$). The results of these fits are shown in Supplementary Figs. 3a-g. Under this constraint, the resistivity tensor is well described at most temperatures; however, noticeable deviations emerge below approximately 20 K

At low temperatures, we relaxed the compensation constraint and allowed $n_\mathrm{e}$ and $n_\mathrm{h}$ to vary independently. This unconstrained two-carrier fit reproduces both $\rho_{xx}(B)$ and $\rho_{yx}(B)$ remarkably well. The mobilities obtained from the two-carrier analysis with and without the compensated metal assumption, as well as those extracted from a single carrier fit, used to define the Hall coefficient in Fig.5d of the main text, are summarized in Supplementary Figs. 3h-i. For reference, in the room temperature crystal structure, a carrier density of one electron per Os corresponds to $12/V_\mathrm{cell} \sim 1.1\times10^{22}$ cm$^{-3}$. Given the semimetallic nature of the compound, the actual carrier density is expected to be be much smaller. The nearly vanishing Hall resistivity above 100 K is therefore naturally interpreted as a signature of a compensated metal with $n_\mathrm{h}\mu_\mathrm{h}^2 \sim n_\mathrm{e}\mu_\mathrm{e}^2$.

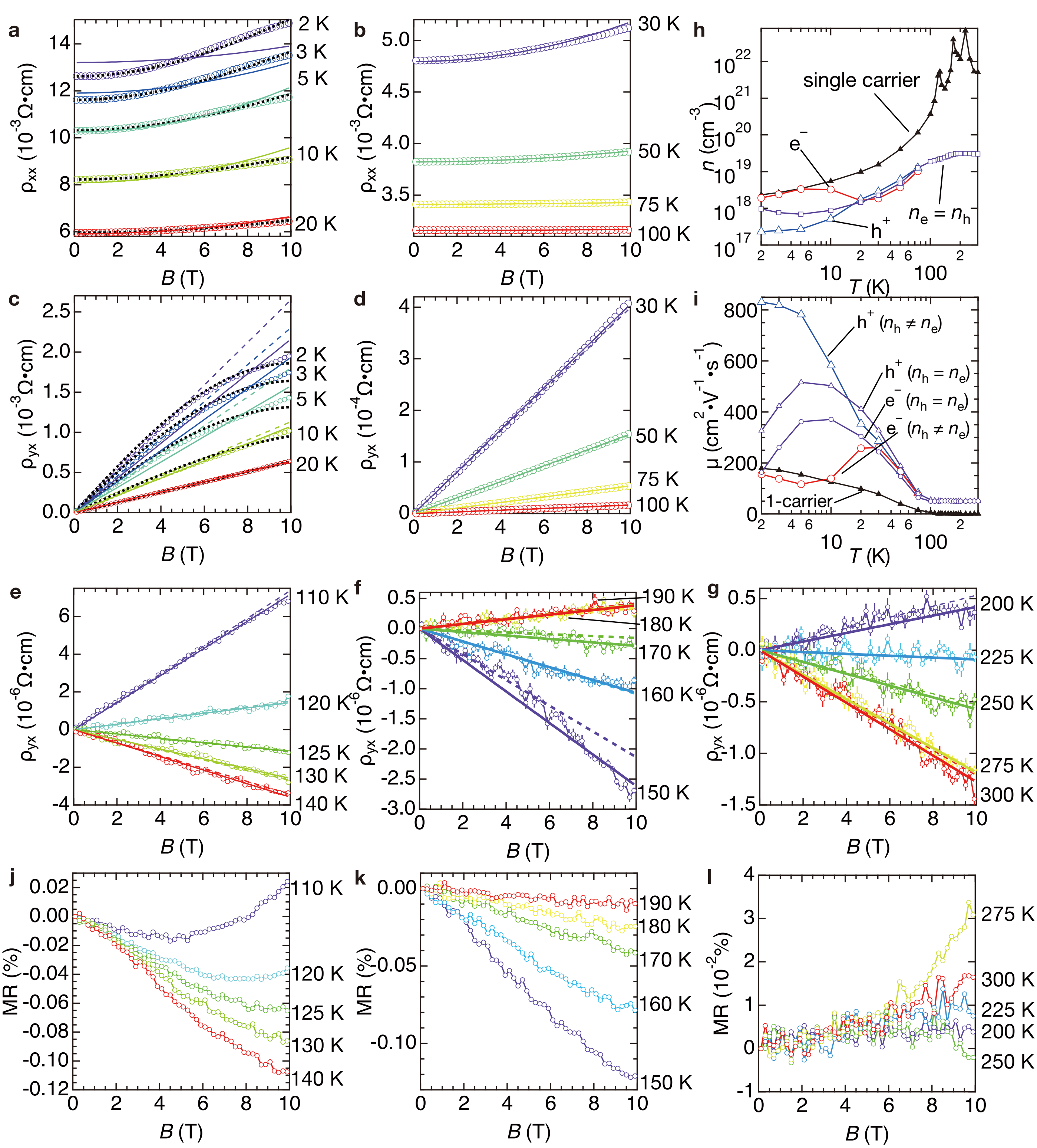

**Supplementary Figure 3 Single-carrier and two-carrier analysis of the magnetotransport.** (a–g) Magnetic-field dependence of the longitudinal $\rho_{xx}$ and transverse $\rho_{yx}$ in-plane resistivities measured between 2 K and 300 K. Experimental data are shown as open circles. Solid colored lines represent fits based on the two-carrier model assuming a compensated metal ($n_e = n_h$), while dotted black lines (a and c) show fits without the compensation constraint ($n_e \neq n_h$). Dashed lines in (c-g) indicate a single carrier fit.

(h, i) Temperature dependence of the carrier density (h) and mobility (i) extracted from single-carrier and two-carrier fits.

(j–l) Temperature dependence of the magnetoresistance above 110 K, exhibiting negative or only weakly field-dependent behavior.

## 6. Band structure calculation

Band-structure calculations were first performed using the monoclinic unit cell to reflect the crystallographic symmetry of $Nd_4Os_3ZnO_{14}$. The result is shown in Supplementary Figure 4.

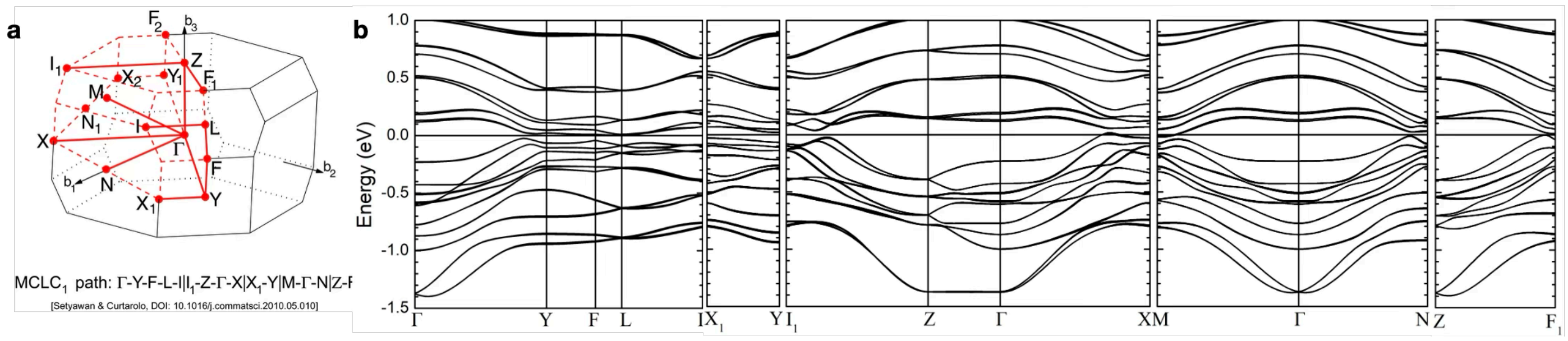

**Supplementary Figure 4**. **The band structure of $Nd_4Os_3ZnO_{14}$ calculated in the monoclinic setting.** (a) Definition of the high-symmetry directions in the monoclinic Brillouin zone. (b) Corresponding band-structure plots along these directions. Although the crystal symmetry is strictly monoclinic, the in-plane lattice parameters are very close to those of a hexagonal lattice as discussed. This allows the electronic structure to be conveniently represented in a pseudo-hexagonal Brillouin zone without altering the qualitative low-energy features. Definition of the special points are Γ(0, 0, 0), $M$(0.249975, 0.41753, 0), and $K$(0.499950, 0.27308, 0). To facilitate comparison with related hexagonal systems and to provide a more intuitive visualization of the band topology, we therefore also present the band structure using a hexagonal representation as shown in Supplementary Figure 5, where band dispersions are plotted for multiple $k_z$ slices in steps of 0.1 $c$*.

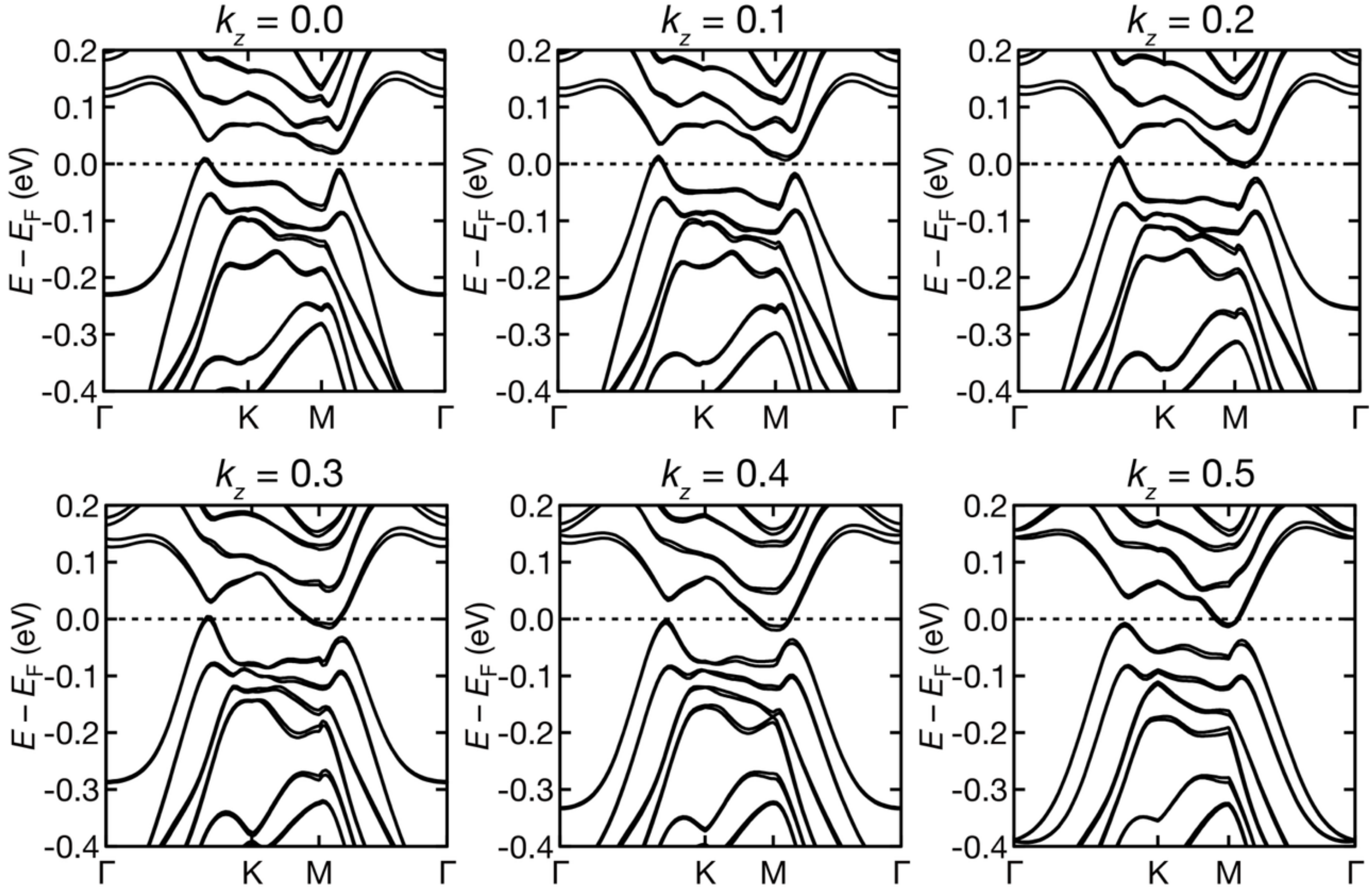

**Supplementary Figure 5**. **Band structure of $Nd_4Os_3ZnO_{14}$ plotted in the hexagonal Brillouin-zone representation.** The definitions of the high-symmetry points are indicated in the figure. In the main text, a schematic cut at $k_z = 0.3c^*$ is used to highlight the coexistence of electron-like and hole-like bands near the Fermi level.